\documentstyle{mn2e}
\input epsf
\def\plotone#1{\centering \leavevmode
\epsfxsize=\columnwidth \epsfbox{#1}}
\def\plottwo#1#2{\centering \leavevmode
\epsfxsize=.99\columnwidth \epsfbox{#1} \hfil
\epsfxsize=.99\columnwidth \epsfbox{#2}}

\newcommand{\beq}{\begin{equation}}
\newcommand{\eeq}{\end{equation}}
\newcommand{\beqa}{\begin{eqnarray}}
\newcommand{\eeqa}{\end{eqnarray}}

\newcommand{\re}{r_{\rm e}}
\newcommand{\mpr}{m_{\rm p}}
\newcommand{\ded}{\Delta E_{\rm D}}

\title[Gas Motions in the Core of the Perseus Cluster]{XMM-Newton Observations
of the Perseus Cluster II: Evidence for Gas Motions in
the Core}

\author[Churazov et al.]{E.~Churazov,$^{1,2}$ W. Forman,$^{3}$
C. Jones,$^{3}$ R.Sunyaev$^{1,2}$ and H. B\"{o}hringer$^{4}$\\
$^1$ Max-Planck-Institut f\"ur Astrophysik, Karl-Schwarzschild-Strasse 1, 85741
Garching, Germany\\
$^2$ Space Research Institute (IKI), Profsoyuznaya 84/32, Moscow 117810, 
Russia\\
$^3$ Harvard-Smithsonian Center for Astrophysics, 60 Garden St.,
Cambridge, MA 02138, USA \\
$^4$ MPI f\"{u}r Extraterrestrische Physik, P.O. Box 1603, 85740
Garching, Germany
}

\pagerange{\pageref{firstpage}--\pageref{lastpage}}
\pubyear{2003}

\begin{document}
\maketitle

\label{firstpage}
\begin{abstract}
The 5-9 keV spectrum of the inner $\sim$100 kpc of the Perseus cluster measured
by XMM-Newton can be well described by an optically thin plasma
emission model as predicted by the APEC code, without any need for
invoking a strong Ni overabundance or the effects of resonant
scattering. For the strongest 6.7 keV line of He-like iron, the optical
depth of the cluster, calculated using observed density, temperature
and abundance profiles, is of order 3. The lack of evidence for
resonant scattering effects implies gas motion in the core with a
range in velocities of at least half of the sound velocity. If this
motion has the character of small scale turbulence, then its
dissipation would provide enough energy to compensate for radiative
cooling of the gas. The activity of the supermassive black hole at
the center of the cluster may be the driving force of the gas
motion.
\end{abstract}

\begin{keywords}
clusters: individual: Perseus - cooling flows
\end{keywords}

%
%________________________________________________________________

\sloppypar

\section{Introduction}
The X-ray emission of the hot gas in galaxy clusters is usually
modeled as emission by an optically thin plasma. The assumption
that the cluster gas is optically thin is certainly valid for the
continuum emission, but for the strongest resonant lines the cluster
can be moderately thick (e.g. Gilfanov, Sunyaev \& Churazov,
1987). Resonant scattering causes changes in line intensities
relative to the continuum, thus affecting measurements of the
heavy element abundances (Gilfanov et al. 1987, Shigeyama 1998), and
producing polarization of the line flux at the level of $\sim$10\%
(Sazonov, Churazov \& Sunyaev, 2002). For a typical rich cluster one
can expect the resonant scattering to be especially important for the
He-like iron $K_\alpha$ line at 6.7 keV. It is most convenient to
search for resonant scattering effects by comparing the flux from
this line with the flux of the He-like iron $K_\beta$ line at 7.9
keV. In particular, for the Perseus cluster, Molendi et al. (1998)
and Akimoto et al. (1997,1999) argued that an anomalously high ratio of
the He-like iron $K_\beta$ and $K_\alpha$ lines hints at the importance of
resonant scattering. The 7.9 keV line of iron is however blended
with the He-like nickel $K_\alpha$ line and the anomalous line ratio
may be interpreted as evidence for an enhanced Ni abundance (Dupke
\& Arnaud, 2001). The role of resonant scattering also has been 
discussed for other objects, in particular M87
(B\"ohringer et al. 2001, Mathews, Buote and Brighenti 2001,
Matsushita, Finoguenov \& B\"ohringer 2003) and NGC~4636 (Xu et al.,
2002).

Below we use an XMM-Newton 50 ksec observation of the Perseus cluster to
assess the possible role of resonant scattering. The detailed
description of the data and the analysis procedure are given in Churazov
et al. (2003).

The structure of the paper is as follows. In Section 2 we simulate the
effect of resonant scattering on the radial profiles of the line
intensities. In Section 3 we compare the results of the observations
with the simulations. In Section 4 we set a lower limit on the level
of gas turbulence, which in the context of this paper means differential 
gas flows on scales smaller than $\sim$100 kpc. 
 In Section 5 we briefly argue that the central
abundance decrement may not be a strong argument in favor of resonant
scattering. Implications of turbulent motions on the thermal balance
of the gas are discussed in Section 6. The last section summarizes our
findings.

Throughout the paper we use $H_0=70~km/s/Mpc$. 

\section{Resonant scattering}
The radial dependence of the electron density $n_e$ and temperature
$T_e$ of the
gas in the Perseus cluster is taken from Churazov et al. (2003),
rescaled to $H_0=70~km/s/Mpc$, namely 
\begin{eqnarray}
n_e=\frac{4.6\times10^{-2}}{[1+(\frac{r}{57})^2]^{1.8}}+
\frac{4.8\times10^{-3}}{[1+(\frac{r}{200})^2]^{0.87}}~~~{\rm cm}^{-3}
\label{ne}
\end{eqnarray}
and
\begin{eqnarray}
T_e=7\frac{[1+(\frac{r}{71})^3]}{[2.3+(\frac{r}{71})^3]}~~~{\rm keV},
\label{te}
\end{eqnarray}
where $r$ is measured in kpc.

\begin{figure} 
\plotone{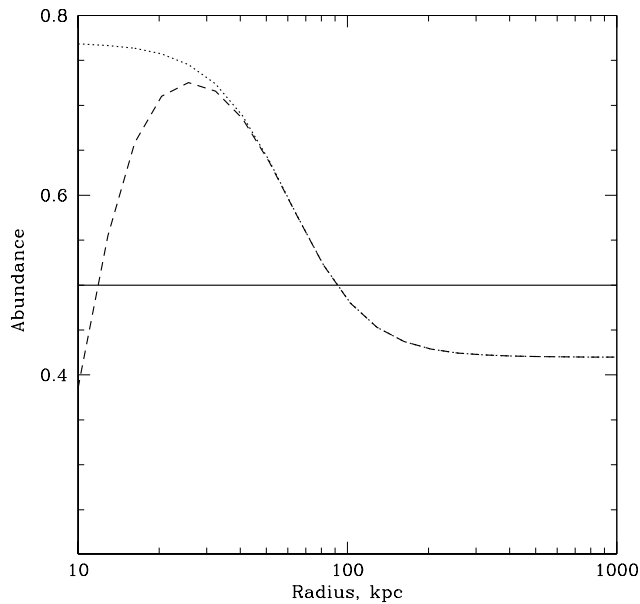}
\caption{Three radial abundance profiles used in the simulations. 
\label{fig:abprof}
}
\end{figure}
For abundances we consider three possible radial behaviors -- a)
constant abundance, b) abundance declining with radius and c)
abundance peaking at the radius of $\sim$50 kpc and declining both
towards smaller or larger radii as shown in Fig.~\ref{fig:abprof}. The
last functional form is closest to the radial abundance profile
derived from deprojection analysis of the Chandra and XMM-Newton
observations (Schmidt, Fabian \& Sanders 2003, Churazov et al. 2003)
under the assumption of a single temperature optically thin plasma
emission model. The abundance ratios used are those of Anders \&
Grevesse (1989).  Using these data we calculated an optical depth from
the center of the cluster up to a radius of 1 Mpc $\tau=\int n_i \sigma_0
dr$, where $n_i$ is the ion concentration and the cross section for a
given ion is \beq \sigma_0=\frac{\sqrt{\pi}h\re c f}{\ded},
\label{sigma_0}
\eeq
where
\beqa
\ded&=& E_0\left(\frac{2kT_e}{A\mpr c^2}+\frac{
V_{\rm turb}^2}{c^2}\right)^{1/2}
\nonumber\\
&=& E_0\left[\frac{2kT_e}{A\mpr c^2}(1+1.4AM^2)\right]^{1/2}.
\label{dop}
\eeqa 
In the above equations $E_0$ is the energy of a given line, $A$
is the atomic mass of the corresponding element, $\mpr$ is the proton
mass, $V_{\rm turb}$ is the characteristic turbulent velocity, $M$
is the corresponding Mach number, $\re$ is the classical electron
radius and $f$ is the oscillator strength of a given atomic
transition. The wavelengths and absorption oscillator strengths are
taken from the compilation of Verner, Verner \& Ferland (1996). The
ionization equilibrium is that of Mazzotta et al. (1998). The set of 
Fe lines with an optical depth larger than 0.2 (for $M=0$)
is given in Table \ref{tab:tau}. From this table it is clear that i)
the 6.7 keV line of He-like iron is by far the most optically thick
line for the case of pure thermal broadening and ii) strong turbulence
makes  resonant scattering effects negligible for all lines as was
emphasized by Gilfanov et al. (1987). For other types of abundance
profiles, the main result is the same -- the 6.7 keV line has an
optical depth of order  3 and accounting for turbulence reduces the
optical depth to values smaller than 1.

\begin{table}
\caption{Optical depth (from $r=0$ to $r=1$ Mpc) to resonant
scattering for a set of X-ray lines.  The temperature and density
profiles are give by eq.~\ref{ne} and \ref{te} above. The abundance is constant
with radius at 0.5 solar. Two values are quoted -- for pure thermal
line broadening ($M=0$) and for strongly turbulent gas ($M=1$). Only
the lines with an optical depth greater than 0.2 (for $M=0$) are
listed.}
\begin{center}
\begin{tabular}{lccl}\hline \\
Ion & Energy & $\tau$ & $\tau$ \\
    & (keV)  & $M=0$  & $M=1$   \\
\\
\hline
Fe XXIII & 0.093 &0.70 & 0.08 \\
Fe XXIII & 1.129 &0.26 & 0.03 \\
Fe XXIV  & 1.163 &0.39 & 0.04 \\
Fe XXIV  & 1.168 &0.78 & 0.09 \\
Fe XXV   & 6.700 &2.79 & 0.31 \\
Fe XXVI  & 6.973 &0.20 & 0.02 \\
Fe XXV   & 7.881 &0.46 & 0.05 \\
\hline
\end{tabular}
\end{center}
\label{tab:tau}
\end{table}

\begin{figure} 
\plotone{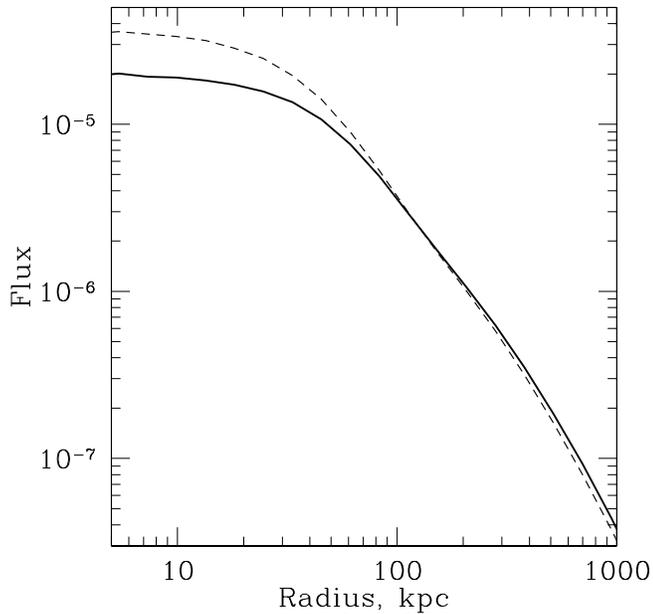}
\caption{Radial profiles of the He-like iron $K_\alpha$ line (for the
flat abundance profile) with (thick solid line) and without (dashed
line) the effect of resonant
scattering. Resonant scattering suppresses the line intensity in the
core and redistributes line photons to larger radii.
\label{fig:lprof}
}
\end{figure}

\begin{figure} 
\plotone{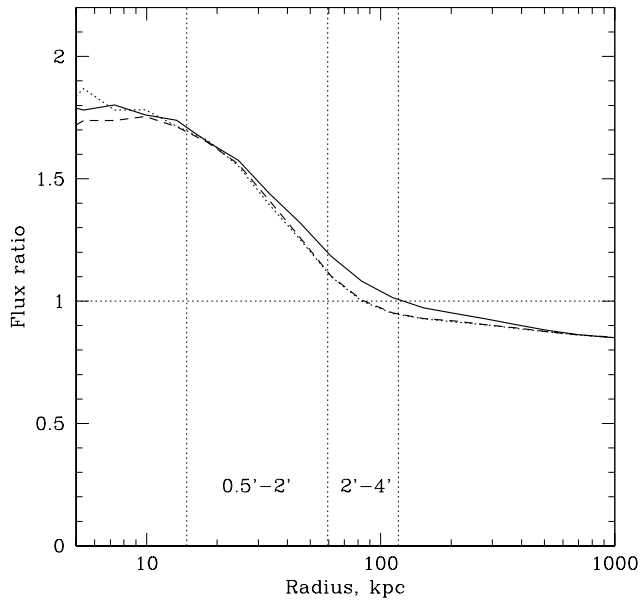}
\caption{Ratio of the 6.7 keV radial brightness profiles without
accounting for resonant scattering to the profiles including the
effect of
resonant scattering. Three curves correspond to three models of the
abundance profiles shown in Fig.~1. Dotted vertical lines show the two
regions used for spectra extraction.
\label{fig:rat}
}
\end{figure}

The resonant scattering has been modeled using a Monte-Carlo
approach. The cluster has been divided into concentric shells and line
emissivities have been assigned to each shell using the APEC v1.3.0
results (Smith et al. 2001). The scattering of photons was accounted
for by assuming a complete energy redistribution and dipole scattering
phase matrix. The latter assumption is motivated by the fact that we
are interested primarily in the He-like iron resonant line, which has
a pure dipole scattering phase matrix (see discussion in Sazonov et
al. 2002). The escaping photons are accumulated into separate bins,
according to their projected distance from the cluster center. The
modification of the 6.7 keV line radial brightness profile due to
resonant scattering (for the case of a flat abundance profile) is
shown in Fig.~\ref{fig:lprof}. As expected, resonant scattering
suppresses the line intensity in the core and raises the line
intensity in the cluster outskirts. More illustrative is the ratio of
the profiles with and without resonant scattering taken into account,
which we show in Fig.~\ref{fig:rat} for all three abundance profile
models. Inspite of the different abundance distributions, this plot
shows that the net effect is very similar for all three cases -- the
flux in the line is suppressed by a factor of up to $\sim$2 within the
inner 100 kpc region and is enhanced by $\sim$10-20\% outside this
region.

\section{Spectra}
\begin{figure} 
\plotone{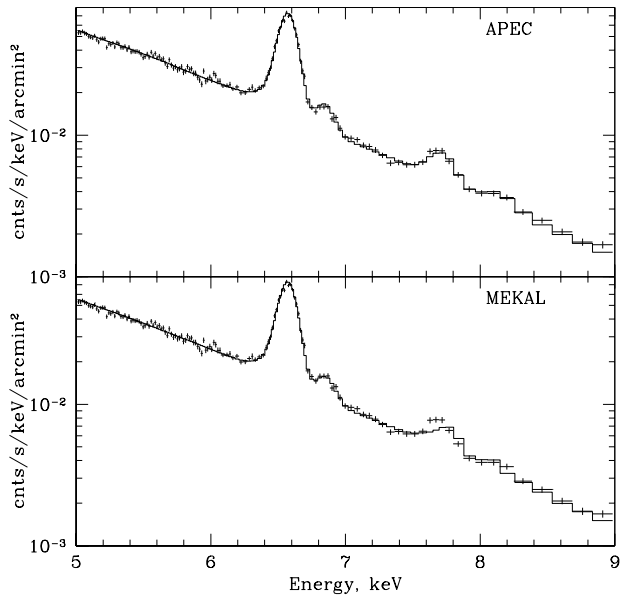}
\caption{The 5-9 keV spectrum of the $30''$ to
$2'$ annulus centered on NGC~1275 and fitted with APEC and MEKAL
models. 
\label{fig:apmek}
}
\end{figure}

Obviously the easiest way to reveal the effect of resonant scattering is to
derive the line ratios for the central region. The He-like iron
$K_\alpha$ and $K_\beta$ lines are ideally suited for that purpose
since both lines are due to the same ion of iron. Guided by
Fig.~\ref{fig:rat} we accumulated MOS spectra for two annuli $0.5'-2'$
and $2'-4'$ centered on NGC~1275. The inner $0.5'$ region was excluded in
order to avoid possible contamination from the NGC~1275 nucleus. The spectrum
from 5 to 9 keV was fitted with the APEC (Smith et al. 2001) and MEKAL
\cite{mewetal85,mewetal86,kaastra92,lieetal95} models in XSPEC
v.11.2 (Arnaud 1996) and is shown in
Fig.~\ref{fig:apmek}. Temperature, heavy metal abundances (with the
abundance ratios of Anders \& Grevesse,
1989) and redshift were free parameters of the models. One can
see that for the MEKAL model there is 
a clear excess on the left wing of the He-like iron $K_\beta$ line, where
the He-like nickel $K_\alpha$ line(s) is blended with the iron line. In
order to remove this discrepancy one has to either raise the nickel
abundance above the standard Ni to Fe ratio (Anders \& Grevesse,
1989) by a factor $\sim$2 or  assume that the 6.7 keV complex is
suppressed by resonant scattering, thus causing peculiarities in the
observed line ratio. On the other hand, the most recent APEC v1.3.0 model
provides an almost perfect fit to the whole 5-9 keV portion of the
spectrum. A very similar situation is seen for the spectrum of the $2'-4'$
annulus.  In what follows we assume that we can rely on the
predictions of the newest APEC model, which has a significantly 
richer set of lines in the region of interest. The best fit
temperature and abundance values for the two annuli are
$T_e=4.29\pm0.05$ keV, $Z=0.501\pm0.007$ relative to the solar
abundance (Anders \& Grevesse, 1989) and $T_e=5.18\pm 0.06$,
$Z=0.446\pm 0.007$. The values of 
temperature are somewhat higher  than one would infer from fitting  a
broader 0.5-9 keV spectral band, which is expected for projected
spectra, given the radial dependence of the temperature in the Perseus
cluster (see eq.~\ref{te}).

The best fit value of the abundance for the 5-9 keV spectrum is of
course dominated by the contribution of the 6.7 keV line.  We  fixed
all parameters (except abundance) at their best fit values and
recalculated the abundance first ignoring part of the spectrum
containing the 6.7 keV complex and second ignoring the part containing 
7.9 keV complex. Since only 6.7 keV
complex is affected by resonant scattering the different values of abundance
in two fits would indicate an important role for scattering.
The ratio of abundances calculated
this way is shown in Fig.~\ref{fig:lmach} with two crosses (for two
annuli). Resonant scattering is expected to increase the ratio of
abundances well above unity for the inner annulus. However, the
observed ratio is consistent with unity, indicating that any effects
from resonant scattering are small.

Finally we demonstrated that projection effects do not significantly
affect the determination of the spectral parameters by using the outer
$2'-4'$ annulus as a background for the inner $0.5'-2'$ annulus. The
resulting spectrum has a somewhat lower best fit temperature $\sim
3.8$ keV, as one would expect, but the line flux ratio again does not
show any obvious anomalies.

\section{Role of gas motions}
\begin{figure} 
\plotone{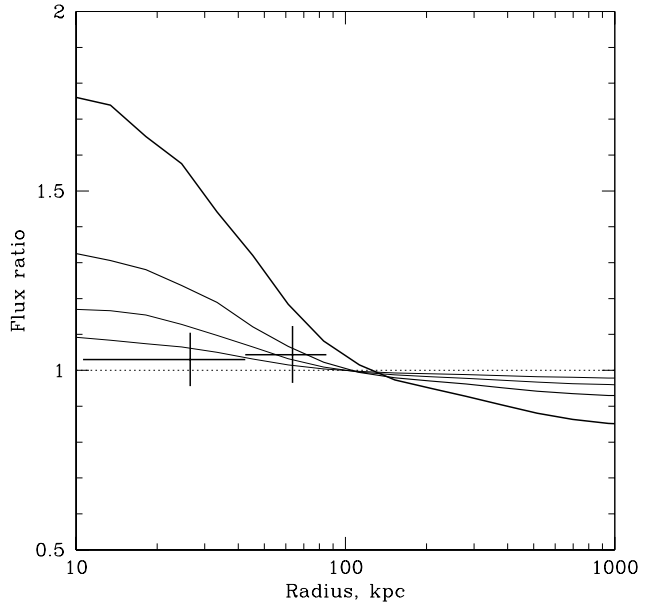}
\caption{Influence of turbulence on the strength of the resonant
scattering effect. The plot shows the same ratio as in
Fig.~\ref{fig:rat} for the flat abundance profile, but calculated for
Mach numbers of 0,0.25,0.5,1 (from top to bottom). For comparison,
crosses show the ratio of heavy elements abundances obtained ignoring
parts of the spectrum containing the 6.7 keV and 7.9 keV complexes
respectively. Given that the resonant $K_\alpha$ line of He-like iron
contributes about 50\% to the 6.6-6.8 keV complex of lines, 
the measured abundance ratio is consistent with the curves for
Mach number $\ge$0.5.
\label{fig:lmach}
}
\end{figure}

The resonant $K_\alpha$ He-like iron line contribution to the 6.7
keV (6.6-6.8 keV) complex varies from
40 to 52\%  for the gas
temperature range from 3 to 5 keV. Therefore suppression of the
resonant line flux in the inner regions due to resonant scattering
would strongly affect the intensity of the whole complex. On the other
hand, the 6.9 and 7.9 keV complexes can be treated as effectively
optically thin.  The good fit of the spectra with the APEC model of an
optically thin plasma with the solar mix of heavy elements can be
considered as an indication that the resonant scattering effects
are suppressed.  As is noted by Gilfanov et al. (1987) turbulent
motions of the gas may significantly reduce the optical depth of the
lines. The effect is especially strong for heavy elements, which have
thermal velocities much smaller than the sound velocity of the
gas. For example, for the 6.7 keV iron line, the inclusion of turbulent motions
(parametrized through the effective Mach number in eq.~\ref{dop}) would
reduce the optical depth to $\sim$0.4 for a Mach number of 1. In
Fig.~\ref{fig:lmach}, we show the simulated ratios of the radial
profiles of the 6.7 keV line with and without the effect of resonant
scattering for Mach numbers of 0, 0.25, 0.5 and 1.

\section{Central abundance hole}

\begin{figure*} 
\plottwo{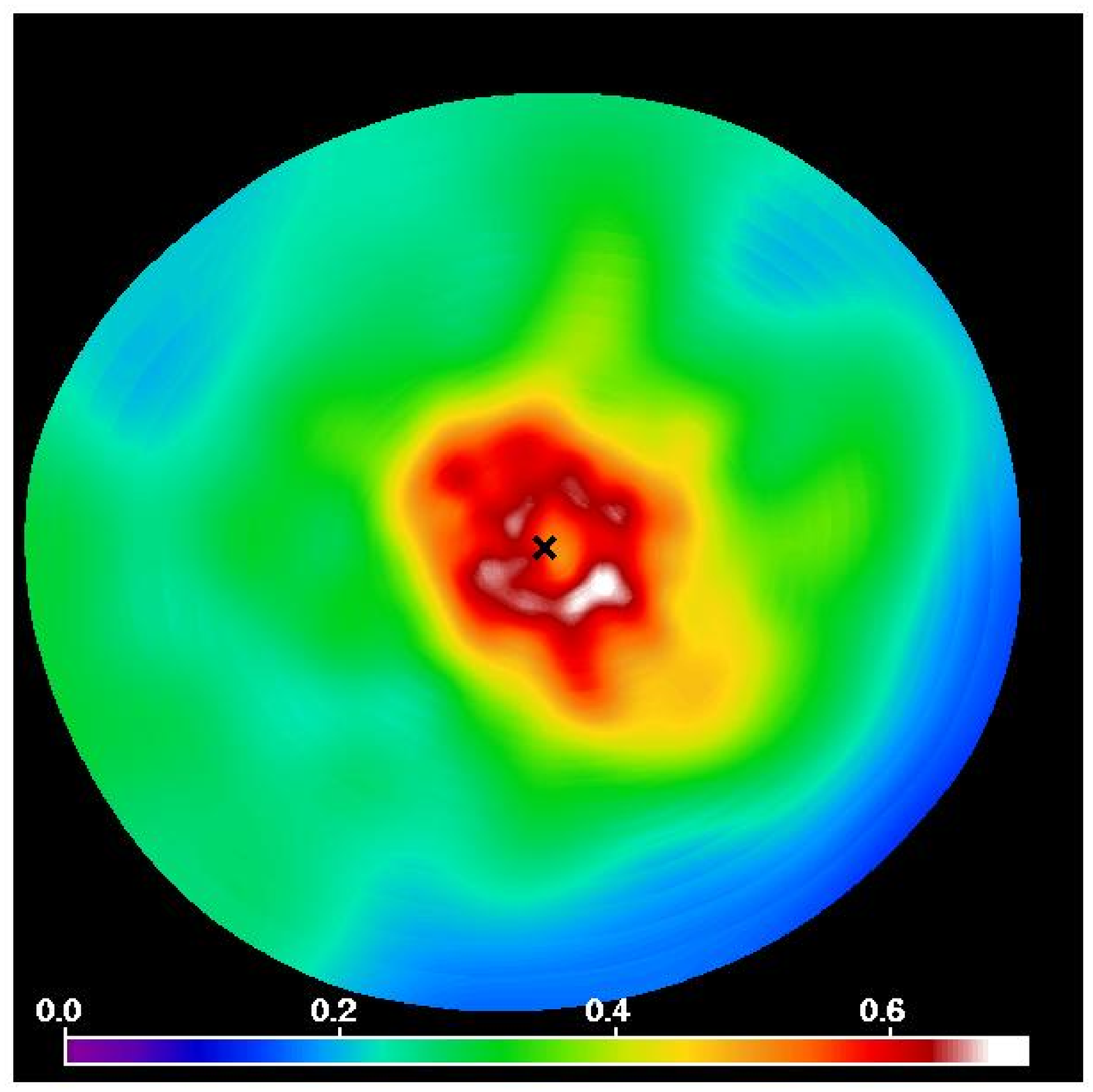}{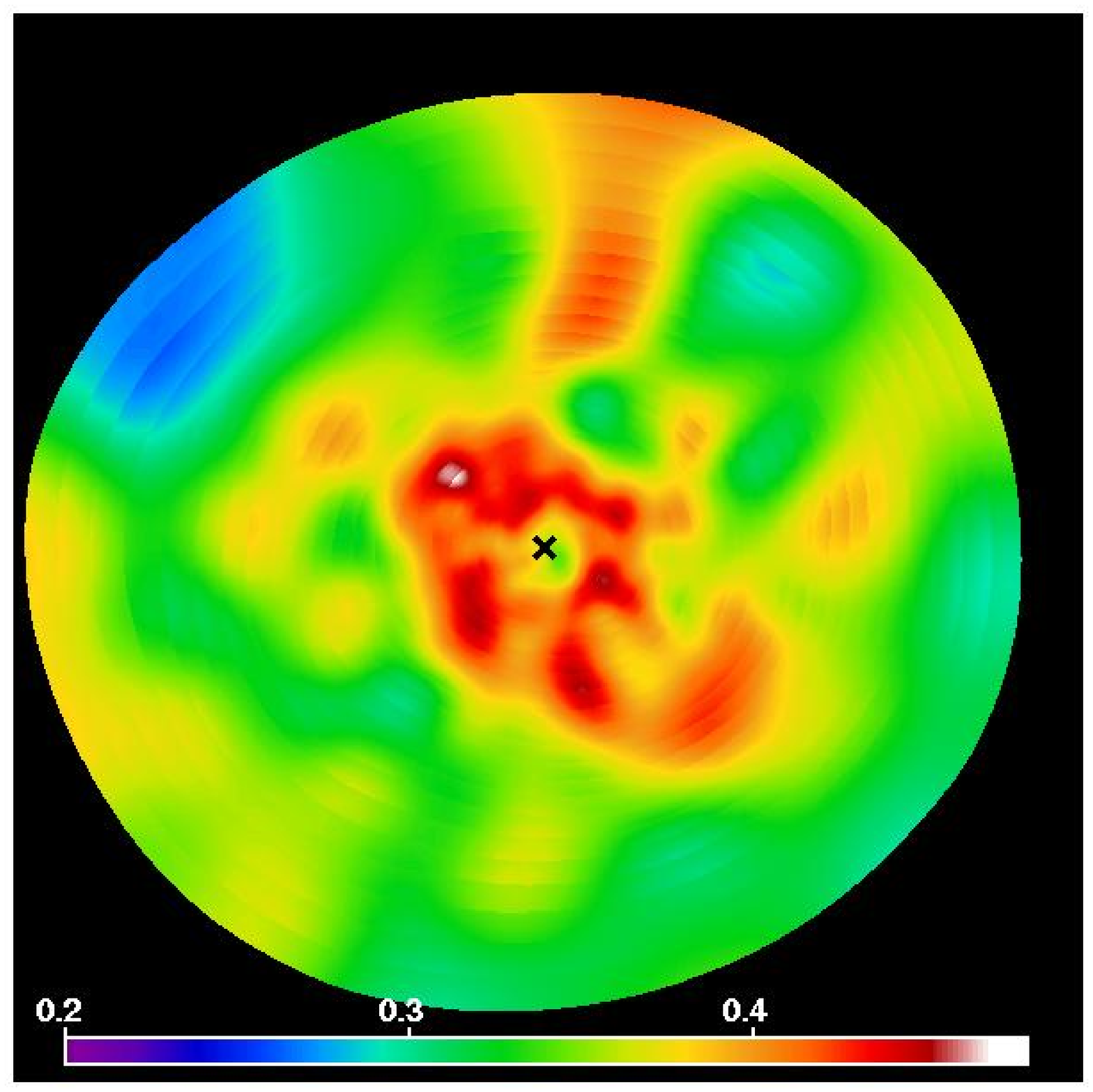}
\caption{{\bf Left:} Adaptively smoothed map of the 6.7 keV line
equivalent width in units of keV. The image size is $20'\times
20'$. Cross marks the position of NGC~1275.  {\bf Right:} The
abundance map (in units of Solar abundance) calculated from the
equivalent width map using the projected temperature map.
\label{fig:ew}
}
\end{figure*}

Single temperature fits to the azimuthly averaged projected spectra of
the Perseus cluster yield an abundance ``hole'' in the very core
(central $\sim 1'$) of the cluster (e.g. Schmidt et al., 2002,
Churazov et al. 2003). It is of course attractive to attribute this
decrease in abundance to resonant scattering, since abundance
measurements are mostly affected by the strongest lines which typically have
the largest optical depths. The high signal-to-noise ratio
accumulated during the XMM-Newton
observations of the Perseus cluster allows one to make a two-dimensional
map of the 6.7 keV line intensity. Shown in Fig.~\ref{fig:ew} (left
panel) is the projected map of the 6.7 keV line equivalent width. The
5-8 keV band of the projected spectra was approximated as a linear
combination of three components - two bremsstrahlung spectra with
temperatures of 2 and 6 keV and a Gaussian line at 6.7 keV (redshifted to the
cluster distance). The very central region (the circle with $20''$
radius centered at NGC~1275) has been excised from the analysis  
to avoid contamination by the AGN flux. The equivalent width of
the line was calculated by applying similar adaptive smoothings to the
intensities of the line and the continuum (sum of two bremsstrahlung
components) and calculating the ratio. The size of the smoothing
window was chosen to provide an effective signal to noise ratio of 80,
calculated using the total number of counts from 5-8 keV. 
The image shows (i) an overall increase in the equivalent
width towards the center of the cluster, (ii) a decrease in the
equivalent width in the central $\sim 1'$ region and (iii) a low equivalent
width horse-shoe shaped region to the West. Since the equivalent width
of the line is temperature dependent, we have calibrated this
dependence using a set of simulated APEC spectra with different
temperatures and a fixed heavy element abundance and applying
a similar procedure for equivalent width determination. Using the
resulting conversion factor (from equivalent width to abundance) and
the temperature map calculated with the same adaptive smoothing
procedure, we converted  the equivalent width map into an abundance
map - Fig.~\ref{fig:ew} (right panel). The low equivalent width region
to the West of the cluster center is apparently the result of higher
gas temperature in this region (see Churazov et al. 2003), since in the
abundance map this feature is absent. The central ``hole'' is present
both in the equivalent width and the abundance map, although detailed
structures are different in the two maps, because of the complicated
temperature structure of the inner region.

\begin{figure*} 
\plottwo{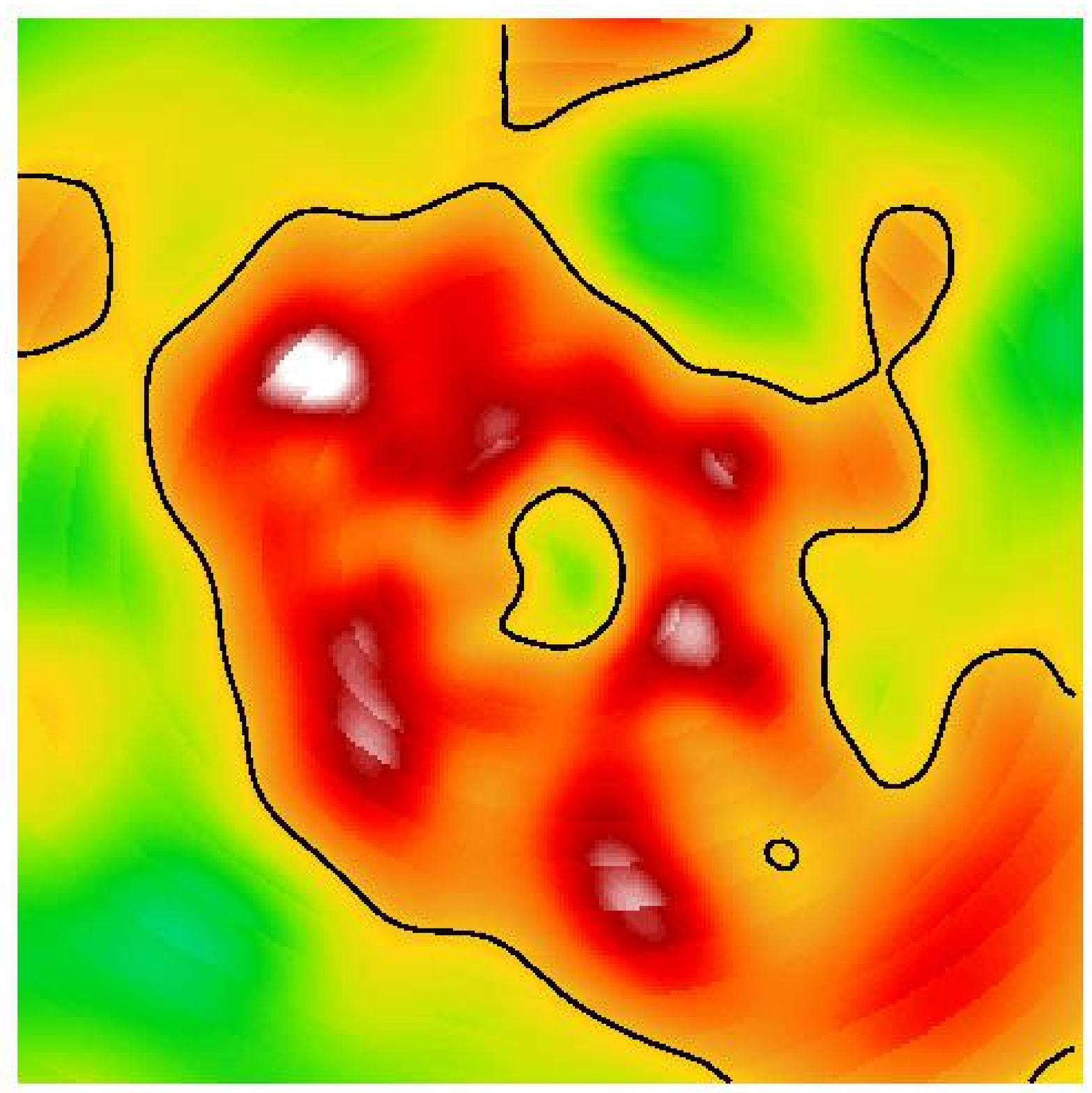}{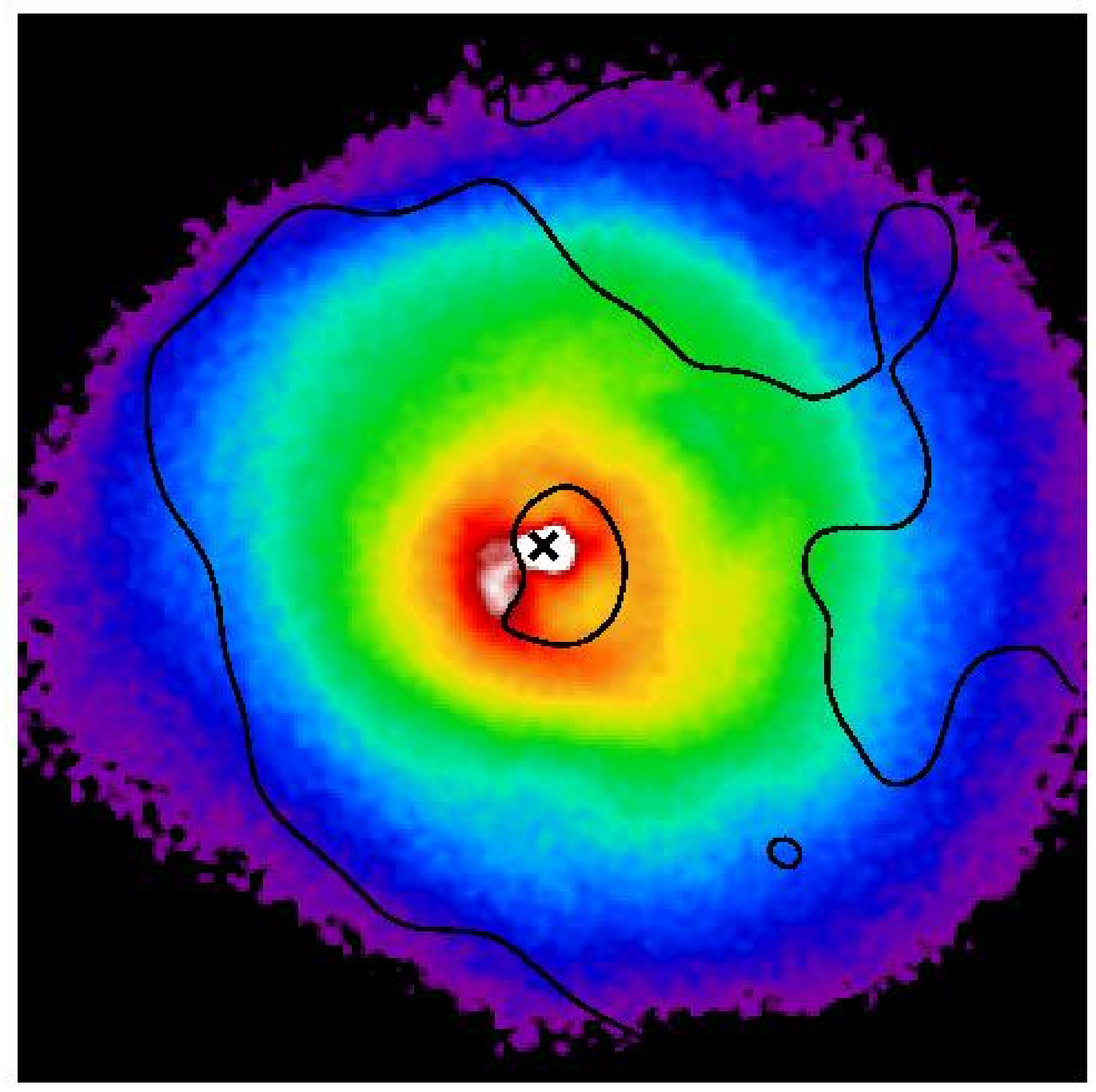}
\caption{{\bf Left:} Central $8'\times 8'$ part of the iron abundance
map. {\bf Right:} Surface brightness distribution of the same region
in the 0.3-5 keV band. Contours in both images show the abundance at the level
of 0.4 Solar. The inner part of the abundance hole approximately covers
two regions of low surface brightness to the North and South of
the nucleus  that correspond to the two radio lobes (B\"ohringer et
al., 1993, Fabian et al. 2000). 
\label{fig:hole2}
}
\end{figure*}

The central part ($8'\times 8'$) of the abundance map is shown in
Fig.~\ref{fig:hole2} along with the surface brightness image of the
same region. Contours in both images show the abundance at the level
of 0.4 Solar. One can see that the inner part of the abundance hole
roughly traces the region occupied by the inner radio lobes
(B\"ohringer et al. 1993, Fabian et al. 2000) which correspond to
areas of low X-ray surface brightness to the North and South of
the nucleus. This morphological similarity suggests that the abundance
``hole'' is related to the activity of the AGN rather than to 
resonant scattering. For instance radio lobes may push the gas with
the highest abundance away from the nucleus forming a shell-like
distribution of highly enriched gas. The details of the abundance
distribution in this region and discussion of realistic models are
beyond the scope of this paper. Given that the size of the region is
rather small and there is a bright AGN at the very center, long Chandra
observations would provide better data for careful study of the
abundance distribution. We conclude however that a central abundance
hole is not a reliable indication that resonant scattering is
important in Perseus and it is likely that other physical mechanisms
play a role.

\section{Discussion}
With our parameterization of the turbulence through the effective Mach
number in eq.~\ref{dop}, the energy in turbulent motions is
approximately related to the thermal energy of the gas as
$\epsilon_{turb}\approx 0.7M^2\epsilon_{th}$. A lack of visible
effects of resonant scattering suggests that $M$ is at least
0.5. Therefore turbulent motions contain at least 20\% of the thermal
energy of the gas. Subsonic turbulence is not very efficient in
generating sound waves (e.g. Landau \& Lifshitz 1963) and one can
assume that a significant fraction of this energy will be dissipated locally
and will go into heat. The dissipation time scale can be estimated as
the 
eddy turn-around time times a factor $f$ of order a few. Thus a rough
estimate of the heating rate due to dissipation of turbulence can
be written as:
\beq
\frac{\epsilon_{turb}}{f\frac{l}{M c_s}}\approx \frac{0.7}{f}M^3
\frac{\epsilon_{th}}{l/c_s}\approx \frac{\epsilon_{th}}{2~10^8~yr},
\label{tdis}
\eeq
where $l$ is the characteristic eddy size, $c_s$ is the sound
velocity. For estimates we set $l=10$ kpc, $M=0.5$ and $f=3$.
This value can be compared with the cooling rate which is set by the thermal
energy and the cooling time which is of the order of 0.5 Gyr for the gas
density and temperature typical of the Perseus core. Thus if
the characteristic spatial scale of turbulent eddies is comparable to or
less than $\sim$10 kpc, then the present level of turbulent dissipation
should be sufficient to compensate for the gas cooling losses. 

There are at least two obvious sources of turbulence -- mergers and
the activity of the central supermassive black hole. The numerical
simulations have shown that mergers can induce long lived (of order of
Gyr) eddies (e.g. Norman \& Bryan 1999) even in cluster cores.  The
surface brightness and gas temperature structures 
in Perseus suggest recent merger activity (e.g. Furusho et
al., 2001). The scale of merger induced eddies is probably rather
large and this reduces the dissipation rate.  On the other hand,
activity of the supermassive black hole can induce turbulence on a
smaller scales through the mechanical action of the outflows (Churazov
et al. 2001,2002, Reynolds, Heinz \& Begelman 2002, Br\"uggen \&
Kaiser 2002). Judging from the size of the observed radio lobes in the
Perseus cluster (B\"ohringer et al. 1993, Fabian et al. 2000) one can
expect the size of the generated eddies to be of order 10 kpc,
i.e. sufficiently small to dissipate their energy quickly (compared to
the timescale for radiative cooling).

The above analysis is simplified by the assumption that the same level
of turbulence, expressed through the effective Mach number, is
applicable to all radii. We cannot prove however with the present data
 that the gas velocities indeed vary on spatial scales as small as
10 kpc, nor is it possible to show that the velocity field can be
characterized as turbulent motion. The conservative conclusion is
therefore that the data are consistent with small scale turbulence
characterized by a velocity of order half the sound speed, although
larger spatial scale velocity variations (up to $\sim$50-100 kpc)
cannot be excluded. Comparable gas velocities are found in the
numerical simulations of cluster formation (e.g. Frenk et al. 1999),
although the central parts of clusters with cool cores may have
properties different from the bulk of the cluster gas. The profile of
a resonant line, broadened by large scale motions and the turbulent
cascade, has been recently calculated by Inogamov \& Sunyaev (2003).
While the profile differs from a simple Gaussian shape, the net effect
on the optical depth in the line core is comparable. We note also that
pure radial differential motions (e.g. sound waves coming from the
very central region as in the picture suggested by Fabian et al. 2003)
also would produce a similar reduction in the optical depth of the
lines. The dissipation rate for this (predominantly radial) motion
crucially depends on the viscosity of the gas (Fabian et
al. 2003). For less regular motion patterns, eq.~\ref{tdis} gives a
resonable estimate of the dissipation time scales for all values of
the viscosity below a certain value (in the limit of large Reynolds 
numbers). For very high viscosities (low Reynolds numbers) the time 
scale for dissipation will be even 
shorter.

Assuming that the gas velocities do vary randomly on spatial
scales of order 10 kpc, one can estimate the impact of these motions
on the radial distribution of heavy elements. In the simplest
approximation one can estimate the turbulent transport of heavy
elements via the diffusion coefficient on spatial scales larger
than the characteristic size of the eddies:
\begin{eqnarray}
D\sim\frac{1}{3}v_t l=\frac{1}{3}M c_s l
\end{eqnarray}
The characteristic time for diffusion over regions of size $X$ 
e.g. 100 kpc is:
\begin{eqnarray}
t\sim \frac{X^2}{D}=~~~~~~~~~~~~~~~~~~~~~~~~~~~~~~~~~~~~~~~~~~~~~~~~~\nonumber \\
6\times10^9 \left ( \frac{X}{100~kpc} \right )^2 \left (
\frac{l}{10~kpc} \right )^{-1} \left (
\frac{M c_s}{500~km/s} \right )^{-1}~yr
\label{td}
\end{eqnarray}
Thus for our choice of typical eddy size and turbulent velocity,
the metals are not transported outside a region much larger than
100 kpc during the lifetime of the cluster. The central abundance
(adopting the abundance profile with a maximum at the center as
shown in Fig.~1), however, does drop substantially on time scales of
approximately 1-2 Gyr and has to be replenished by some mechanism (see
B\"ohringer et al. (2003) for implications of abundance gradients
on the properties of the cool core clusters).  Given the different
dependence of the turbulent diffusion coefficient (eq.\ref{td}) and
the rate of dissipation (eq.\ref{tdis}) on the spatial scales of
eddies and characteristic velocities, the importance of these two
processes will differ.  For some combinations of parameters (in
particular for small eddies), the heating rate is high while the
impact of turbulent transport may be limited. Since at present
these values are highly uncertain, it is difficult to prove that such a
situation is indeed taking place in Perseus.

The are several additional caveats associated with the above analysis
which are necessary to mention. First of all there are still issues to
be resolved in the plasma emission models, in particular near the
He-like Ni $K_\alpha$ line.  Compared to the MEKAL version included
in XSPEC  (version 11.2.0), the APEC code has an updated set of major
line energies and atomic physics (Smith et al. 2001, see also {\bf
http://cxc.harvard.edu/atomdb}). For the MEKAL model, however, the best
description of the spectra are obtained when Ni is overabundant 
(Dupke \& Arnaud, 2001, Gastaldello \& Molendi, 2003) rather
than by effects of  resonant scattering. It seems therefore that even
although the predictions of two codes differ, both
codes favor a minimal role for resonant scattering.
Second, the presence of
multi-temperature plasma in the Perseus core and projection effects
make the straightforward interpretation of simple single temperature
fits to the projected spectra questionable. Finally, the XMM-Newton
observation of the Perseus cluster was affected by increased
background (see Churazov et al. 2003 for details) which might
slightly affect the spectral parameters. We believe however that all
these problems should not affect our estimate that the
random gas velocities in the Perseus cluster core are a
substantial fraction of the gas sound speed.

The presence of differential motions and the role of the
resonance scattering in the Perseus core can be verified in future
X-ray observations. The most straightforward way would be the
measurements of the line width with calorimeters (ASTRO-E2,
Constellation-X, XEUS). Resonant lines are broadened both by the
resonant scattering (e.g. Gilfanov et al., 1987) and the differential
motions (e.g. Inogamov and Sunyaev 2003; Sunyaev, Norman and Bryan
2003), while the width of the forbidden or intercombination lines is
affected only by differential motions. Comparison of the lines width
would provide an important test on the contribution of the resonant
scattering. At the temperature of 4 keV the pure Doppler width of the
iron 6.7 keV resonant line is of order of 4 eV (FWHM) or $\sim$200
km/s. Resonant scattering (for an optical depth of order of 3) would
double the width of the resonant line (Gilfanov et al., 1987). These
values are comparable with the ASTRO-E2 energy resolution and could be
measured. Shape of the line and mapping of the line centroid energy
over the central region will help to identify characteristic size of
the eddies (e.g. Inogamov and Sunyaev 2003).

A more demanding test for the presence of resonant scattering would be
the measurements of the polarization of the line flux (Sazonov et al.,
2002). The expected degree of polarization for the central region is
however small (at the level of few per cents) which makes a detection or
meaningful upper limits problematic for future polarimetric projects.

The third possibility is related to $H_\alpha$ filaments observed in
the core of the Perseus cluster. If the filaments are constructed from
an extremely small clouds as suggested by Fabian et al. 2003 which are
dragged by the flow of the hot gas then one can get a direct estimate
of the gas velocity spread $\sim 300 km/s$ (e.g. Conselice, Gallagher
\& Wyse, 2001) and the
characteristic scales of the eddies ($\sim 10$ kpc) from the optical
data. We note that if individual clouds are not very small,
have a temperature of $10^4$ K and are in pressure equilibrium with
the ambient hot gas at $3~10^7$ K then the density contrast of $c\sim
3000$ will make them insensitive to the varying gas velocity field on
the spatial scales smaller than $\sim c\times x$, where $x$ is the size of the
cloud. For instance if the size of the typical cloud is $\sim 1$
pc, then it has to travel $\sim 3$ kpc through the ambient gas
before ram pressure will accelerate it to the velocity comparable with
the gas velocity.

\section{Conclusions}

The Perseus cluster spectrum from 5 to 9 keV measured by XMM-Newton
can be well described as emission from an optically thin plasma
with solar ratios of elemental abundances, provided that the most
recent version (1.3) of the APEC code is used.

The lack of any visible suppression of the He-like iron 6.7 keV line
in the inner region of the Perseus cluster core implies that the
optical depth of the cluster gas is significantly reduced by gas
motions with velocities of order half the gas sound
velocity. Dissipation of these motions may provide enough heat to
replenish the energy lost from radiative cooling if the spatial
scales of velocity variations are small enough (e.g. comparable to the
size of AGN-inflated relativistic plasma bubbles).

\section{Acknowledgements} 

We are grateful to Randall Smith for useful discussions. We thank the editor 
and the referee for important comments and suggestions. W. Forman and
C. Jones thank MPA for its hospitality during their summer 2002
visits, as well as the Smithsonian Institution and Chandra Observatory
for support (NASA contract NAS8-39073).  This work is based on
observations obtained with XMM-Newton, an ESA science mission with
instruments and contributions directly funded by ESA Member States and
the USA (NASA).

\label{lastpage}


\begin{thebibliography}{}

\bibitem[\protect\citename{Akimoto et al. }1997]{akietal97} Akimoto
F. et al., 1997, in X-ray Imaging and Spectroscopy of Cosmic Plasmas,
ed. F. Makino \& K. Mitsuda (Tokio: Universal Academy Press), 95

\bibitem[\protect\citename{Akimoto} 1999]{1999AN....320..283A} Akimoto F., 
Furuzawa A., Tawara Y., Yamashita K., 1999, AN,  320, 283 


\bibitem[\protect\citename{Anders} 1989]{1989GeCoA..53..197A} Anders E., 
Grevesse N., 1989, GeCoA,  53, 197 

\bibitem[Arnaud (1996)]{ar96} Arnaud K.A. 1996,
         Astronomical Data Analysis Software and Systems V,
         eds. Jacoby G. and Barnes J., p17, ASP Conf. Series volume
         101.  


\bibitem[\protect\citename{B\"ohringer} 1993]{1993MNRAS.264L..25B} 
B\"ohringer H., Voges W., Fabian A.~C., Edge A.~C., Neumann D.~M., 1993, 
MNRAS,  264, L25 

\bibitem[\protect\citename{B\"ohringer} 2003]{2003b} 
B\"ohringer H. et al., 2003, A\&A, to be submitted.

\bibitem[\protect\citename{B{\" o}hringer} 2001]{2001A&A...365L.181B} B{\" 
o}hringer H., Belsole E., Kennea J., et al., 2001, A\&A,  365, L181 

\bibitem[\protect\citename{Br{\" u}ggen} 2002]{2002Natur.418..301B} Br{\" 
u}ggen M., Kaiser C.~R., 2002, Nature,  418, 301 

\bibitem[\protect\citename{Churazov} 2001]{2001ApJ...554..261C} Churazov 
E., Br{\" u}ggen M., Kaiser C.~R., B{\" o}hringer H., Forman W., 2001, ApJ, 
554, 261 

\bibitem[\protect\citename{Churazov} 2002]{2002MNRAS.332..729C} Churazov 
E., Sunyaev R., Forman W., B{\" o}hringer H., 2002, MNRAS,  332, 729 


\bibitem[\protect\citename{Churazov} 2003]{2003ApJ...xxxx..261C}
Churazov E., Forman W., Jones C., B{\" o}hringer H., 2003, ApJ, 590,
225.

\bibitem[Conselice, Gallagher, \& Wyse(2001)]{2001AJ....122.2281C} 
Conselice, C.~J., Gallagher, J.~S., \& Wyse, R.~F.~G.\ 2001, AJ, 122,
2281 

\bibitem[\protect\citename{Dupke} 2001]{2001ApJ...548..141D} Dupke R.~A., 
Arnaud K.~A., 2001, ApJ,  548, 141 

\bibitem[\protect\citename{Fabian} 2000]{2000MNRAS.318L..65F} Fabian A.~C., 
Sanders J.~S., Ettori S., et al., 2000, MNRAS,  318, L65 

\bibitem[\protect\citename{Fabian} 2003]{2003MNRAS.318L..65F} Fabian
A.~C. et al., 2003, MNRAS, accepted 

\bibitem[\protect\citename{Frenk} 1999]{1999ApJ...525..554F} Frenk C.~S., 
White S.~D.~M., Bode P., et al., 1999, ApJ,  525, 554 

\bibitem[\protect\citename{Furusho} 2001]{2001ApJ...561L.165F} Furusho T., 
Yamasaki N.~Y., Ohashi T., Shibata R., Ezawa H., 2001, ApJ,  561, L165 

\bibitem[gm03]{gm03} Gastaldello F. \& Molendi S., 2003, talk at the
Conference ``The Riddle of Cooling Flows in Galaxies and Clusters of
Galaxies'', Charlottesville, May 31 -- June 4,
http://www.astro.virginia.edu/coolflow/abs.php 



\bibitem[\protect\citename{Gilfanov} 1987]{1987SvAL...13..233G} Gilfanov 
M.~R., Sunyaev R.~A., Churazov E.~M., 1987, SvAL,  13, 233 

\bibitem[\protect\citename{Inogamov} 2003]{2003SvAL...13..233G}
Inogamov N., Sunyaev R., 2003, Astronomy Letters, in press

\bibitem[\protect\citename{Kaastra }1992]{kaastra92} Kaastra J.S.,
1992, An X-Ray Spectral Code for Optically Thin Plasmas (Internal
SRON-Leiden Report, updated version 2.0)

\bibitem{ll63} Landau, L.D., Lifshitz, E.M., 1963, Fluid mechanics,
Pergamon Press 

\bibitem[\protect\citename{Liedahl et al. }1995]{lieetal95} Liedahl D.A.,
Osterheld A.L. and Goldstein W.H., 1995, ApJL, 438, 115

\bibitem[\protect\citename{Mathews} 2001]{2001ApJ...550L..31M} Mathews 
W.~G., Buote D.~A., Brighenti F., 2001, ApJ,  550, L31 

\bibitem[\protect\citename{Matsushita} 2003]{2003A&A...401..443M} 
Matsushita K., Finoguenov A., B{\" o}hringer H., 2003, A\&A,  401, 443 

\bibitem[\protect\citename{Mazzotta} 1998]{1998A&AS..133..403M} Mazzotta 
P., Mazzitelli G., Colafrancesco S., Vittorio N., 1998, A\&AS,  133, 403 


\bibitem[\protect\citename{Mewe et al. }1986]{mewetal85} Mewe R.,
Gronenschild E.H.B.M. and van den Oord G.H.J., 1985, A\&AS, 62, 197 

\bibitem[\protect\citename{Mewe et al. }1986]{mewetal86} Mewe R., Lemen
J.R. and van den Oord G.H.J., 1986, A\&AS, 65, 511 

\bibitem[\protect\citename{Molendi} 1998]{1998ApJ...499..608M} Molendi S., 
Matt G., Antonelli L.~A., Fiore F., Fusco-Femiano R., Kaastra J., Maccarone 
C., Perola C., 1998, ApJ,  499, 608 

\bibitem[\protect\citename{Norman} 1999]{1999rgm87conf..106N} Norman M.~L., 
Bryan G.~L., 1999, in Lecture Notes in Physics 530, The Radio Galaxy
Messier 87, ed. H.-J. R\"oser \& K. Meisenheimer (New York: Springer), 106 

\bibitem[\protect\citename{Reynolds} 2002]{2002MNRAS.332..271R} Reynolds 
C.~S., Heinz S., Begelman M.~C., 2002, MNRAS,  332, 271 


\bibitem[\protect\citename{Sazonov} 2002]{2002MNRAS.333..191S} Sazonov 
S.~Y., Churazov E.~M., Sunyaev R.~A., 2002, MNRAS,  333, 191 

\bibitem[\protect\citename{Schmidt} 2002]{2002MNRAS.337...71S} Schmidt 
R.~W., Fabian A.~C., Sanders J.~S., 2002, MNRAS,  337, 71 

\bibitem[\protect\citename{Shigeyama} 1998]{1998ApJ...497..587S} Shigeyama 
T., 1998, ApJ,  497, 587 

\bibitem[\protect\citename{Smith} 2001]{2001ApJ...556L..91S} Smith R.~K., 
Brickhouse N.~S., Liedahl D.~A., Raymond J.~C., 2001, ApJ,  556, L91 

\bibitem[\protect\citename{Sunyaev} 2003]{2003SvAL...13..233S}
Sunyaev R., Norman M., Bryan G., 2003, Astronomy Letters, in press

\bibitem[\protect\citename{Verner} 1996]{1996ADNDT..64....1V} Verner D.~A., 
Verner E.~M., Ferland G.~J., 1996, ADNDT,  64, 1 


\bibitem[\protect\citename{Xu} 2002]{2002ApJ...579..600X} Xu H., Kahn 
S.~M., Peterson J.~R., et al., 2002, ApJ,  579, 600 

\end{thebibliography}
\end{document}